\documentclass[twocolumn,tighten,times,twocolappendix,numberedappendix]{aastex631}

\usepackage{amsmath}
\usepackage{color}
\usepackage{xcolor}
\usepackage{url}
\usepackage{stmaryrd}
\usepackage{etoolbox}

\usepackage{graphicx}
\usepackage[flushleft]{threeparttable}
\usepackage{placeins}
\usepackage{comment}

\usepackage[encapsulated]{CJK}
\usepackage{ucs}
\usepackage[utf8x]{inputenc}
\newcommand{\cntext}[1]{\begin{CJK}{UTF8}{gbsn}#1\end{CJK}}

\hypersetup{breaklinks=true,colorlinks=true,linkcolor=blue,citecolor=blue,filecolor=blue,urlcolor=blue}
\makeatletter
\patchcmd{\NAT@citex}
  {\@citea\NAT@hyper@{%
     \NAT@nmfmt{\NAT@nm}%
     \hyper@natlinkbreak{\NAT@aysep\NAT@spacechar}{\@citeb\@extra@b@citeb}%
     \NAT@date}}
  {\@citea\NAT@nmfmt{\NAT@nm}%
   \NAT@aysep\NAT@spacechar\NAT@hyper@{\NAT@date}}{}{}

\patchcmd{\NAT@citex}
  {\@citea\NAT@hyper@{%
     \NAT@nmfmt{\NAT@nm}%
     \hyper@natlinkbreak{\NAT@spacechar\NAT@@open\if*#1*\else#1\NAT@spacechar\fi}%
       {\@citeb\@extra@b@citeb}%
     \NAT@date}}
  {\@citea\NAT@nmfmt{\NAT@nm}%
   \NAT@spacechar\NAT@@open\if*#1*\else#1\NAT@spacechar\fi\NAT@hyper@{\NAT@date}}
  {}{}
  
\usepackage{array}
\newcolumntype{C}[1]{>{\centering\let\newline\\\arraybackslash\hspace{0pt}}m{#1}}
\usepackage{xspace}

\def\apj{ApJ}
\def\apjl{ApJ}
\def\apjs{ApJS}
\def\ao{Appl.~Opt.}

\def\aap{A\&A}

\def\mnras{MNRAS}

\def\procspie{Proc.~SPIE}

\def\nar{New~Astro.~Rev.}

\def\arcdeg{\hbox{$^\circ$}}

\definecolor{burgundy}{rgb}{0.5, 0.0, 0.13}

\usepackage{xspace}

\newcommand{\orcidicon}{\includegraphics[width=0.26cm]{orcid-ID.eps}}

\newcommand{\orcidauthor}[1]{\href{https://orcid.org/#1}{\orcidicon}}

\shorttitle{An EPIC X-ray view of R Aqr}
\shortauthors{Toal\'{a} et al.}

\makeatletter
\patchcmd{\frontmatter@RRAP@format}{(}{}{}{}
\patchcmd{\frontmatter@RRAP@format}{)}{}{}{}
\renewcommand\Dated@name{}
\makeatother

\begin{document}

\title{\large An {\it XMM-Newton} EPIC X-ray view of the Symbiotic Star R~Aquarii}

\correspondingauthor{Jes\'{u}s A. Toal\'{a}}
\email{j.toala@irya.unam.mx}

\author[0000-0002-5406-0813]{Jes\'{u}s\,A.\,Toal\'{a}\cntext{(杜宇君)}}
\affil{Instituto de Radioastronom\'ia y Astrof\'isica,  UNAM 
Campus Morelia, Apartado postal 3-72, 58090, Morelia, Michoacán, Mexico}

\author[0000-0003-0242-0044]{Laurence~Sabin}
\affil{Instituto de Astronom\'{i}a, UNAM, Apdo. Postal 877, Ensenada 22860, B.C., Mexico}

\author[0000-0002-7759-106X]{Mart\'{i}n~A.\,Guerrero}
\affil{Instituto de Astrof\'{i}sica de Andaluc\'{i}a, IAA-CSIC, Glorieta de la Astronom\'{i}a S/N, E-18008 Granada, Spain}

\author[0000-0003-2653-4417]{Gerardo~Ramos-Larios}
\affil{Instituto de Astronom\'\i a y Meteorolog\'\i a, CUCEI, 
Univ.\ de Guadalajara, 
Av.\ Vallarta 2602, Arcos Vallarta, 44130 Guadalajara, Mexico}

\author[0000-0003-3667-574X]{You-Hua\,Chu\cntext{(朱有花)}}
\affil{Institute of Astronomy and Astrophysics, Academia Sinica (ASIAA), No. 1, Sec. 4, Roosevelt Road, Taipei 10617, Taiwan}

\date[ ]{Submitted to ApJL}

\begin{abstract}
\noindent 
We present the analysis of archival {\it XMM-Newton} European Photon
Imaging Camera (EPIC) X-ray observations of the symbiotic star
R~Aquarii. We used the Extended Source Analysis Software (ESAS)
package to disclose diffuse soft X-ray emission extending up to
2.2~arcmin ($\approx$0.27~pc) from this binary system. The depth of
these {\it XMM-Newton} EPIC observations reveal in unprecedented
detail the spatial distribution of this diffuse emission, with a
bipolar morphology spatially correlated with the optical nebula. The
extended X-ray emission shares the same dominant soft X-ray-emitting
temperature as the clumps in the jet-like feature resolved by {\it
  Chandra} in the vicinity of the binary system. The harder component
in the jet might suggest that the gas cools down, however, the
possible presence of non-thermal emission produced by the presence of
a magnetic field collimating the mass ejection can not be
discarded. We propose that the ongoing precessing jet creates bipolar
cavities filled with X-ray-emitting hot gas that feeds the more
extended X-ray bubble as they get disrupted. These EPIC observations
demonstrate that the jet feedback mechanism produced by an accreting
disk around an evolved, low-mass star can blow hot bubbles, similar to
those produced by jets arising from the nuclei of active galaxies.
\end{abstract}


\keywords{\href{http://astrothesaurus.org/uat/1674}{Symbiotic binary stars (1674)};
\href{http://astrothesaurus.org/uat/1607}{Stellar jets (1607)};
\href{http://astrothesaurus.org/uat/1636}{Stellar winds (1636)};
\href{http://astrothesaurus.org/uat/1822}{X-ray sources (1822)};
\vspace{4pt}
\newline
}


\section{Introduction}
\label{sec:intro}

R Aquarii (R Aqr) is a symbiotic star (SS) consisting of a 385-day
period pulsating Mira-like star and a hot white dwarf (WD) harboring
an accretion disk.  At its {\it Gaia} geometric distance of
385$\pm$60~pc, it is one of the closest and best
studied SS, with detailed determinations of its orbital parameters
\citep{Hollis1997,Gromadzki2009,Bujarrabal2018}, physical
characteristics \citep{Schmid2017} and dust content
\citep[see][]{Mayer2013}.

\begin{figure*}
\begin{center}
\includegraphics[width=0.5\linewidth]{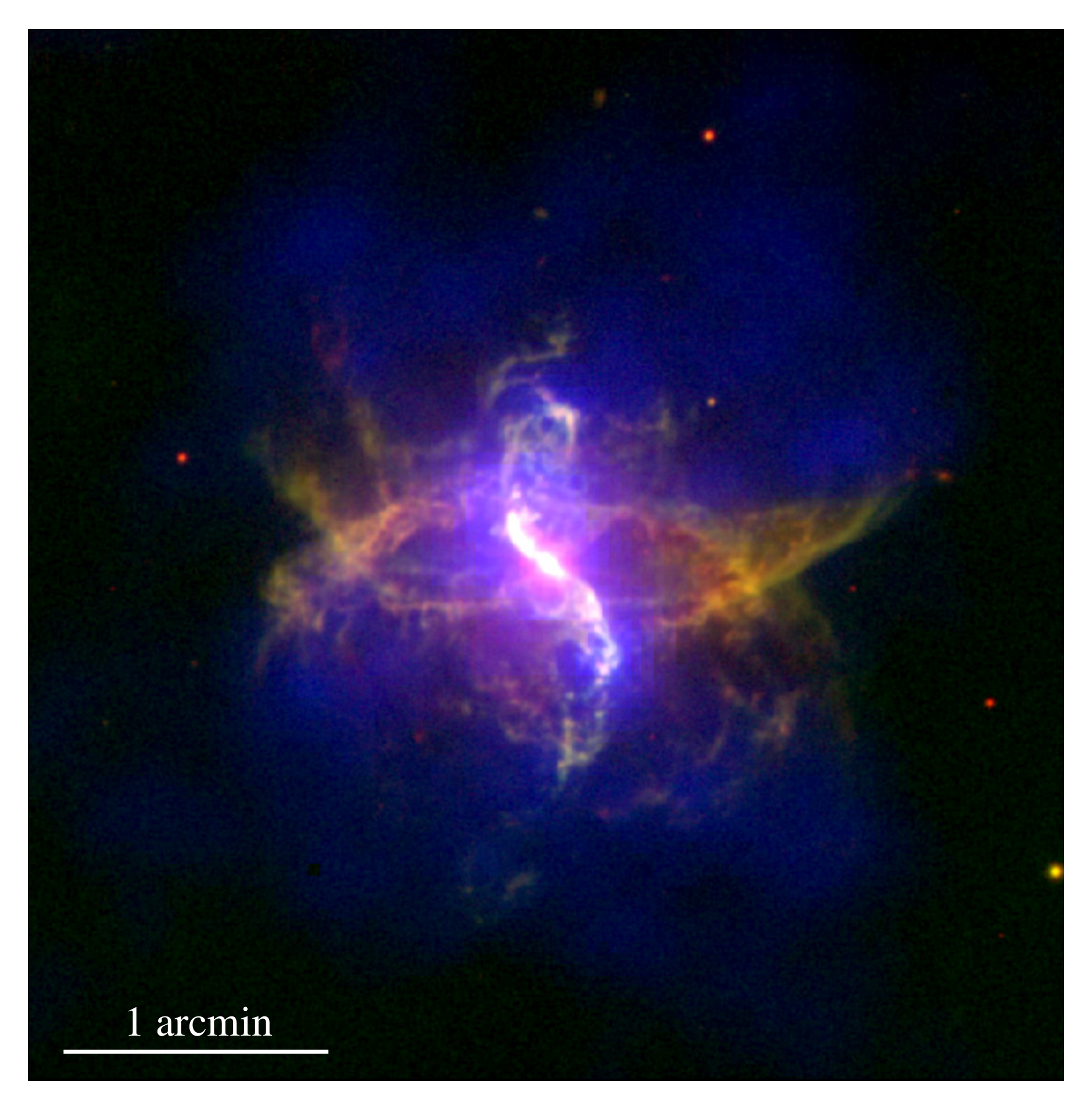}~
\includegraphics[width=0.5\linewidth]{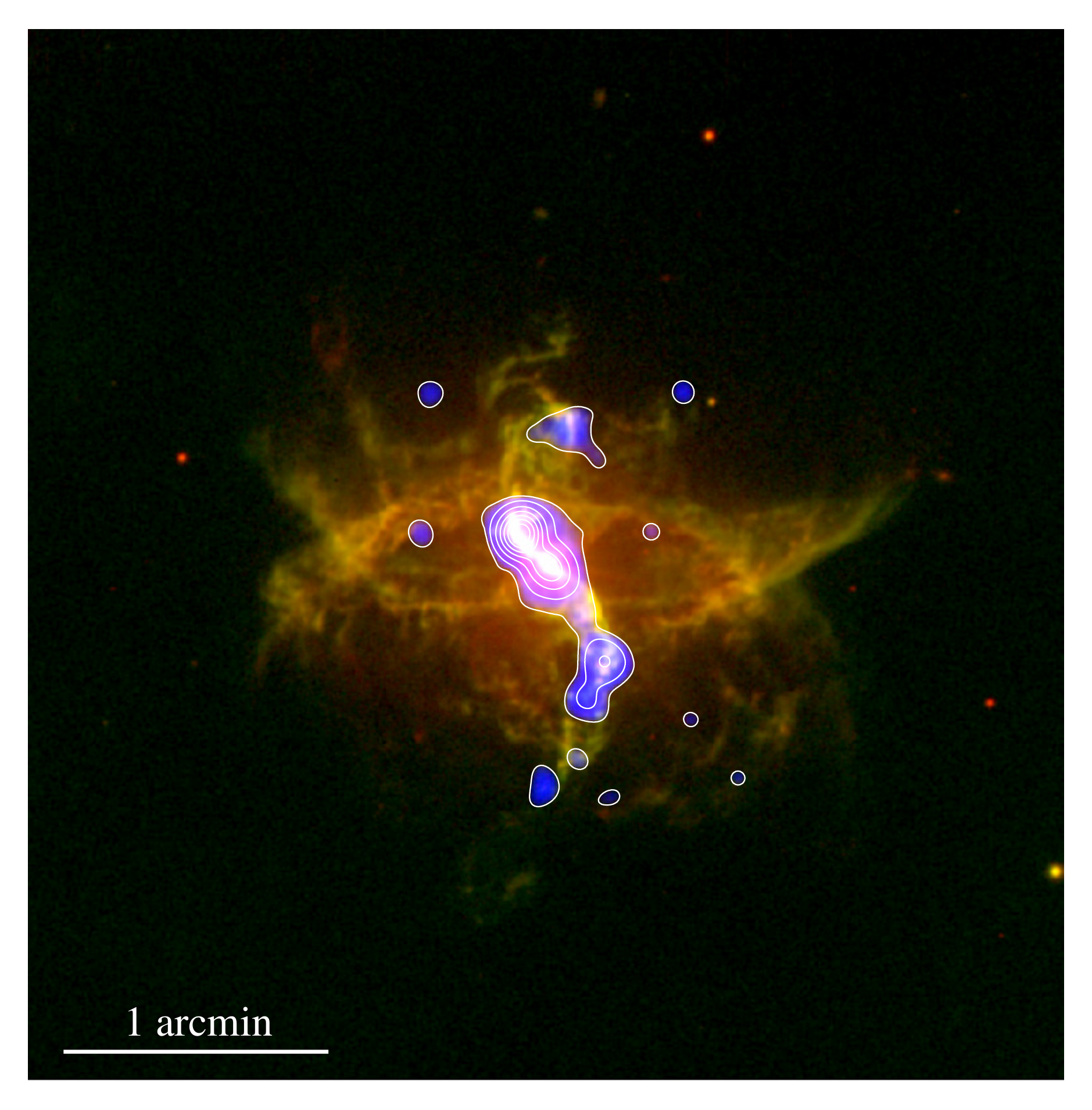}
\caption{Color-composite images of R~Aqr.  The soft (0.3--0.7~keV)
  X-ray emission detected by {\it XMM-Newton} EPIC (left) and {\it
    Chandra} ACIS-S (right) is presented in blue. Green and red in
  both panels correspond to the [O\,{\sc ii}] and H$\alpha$+[N\,{\sc
      ii}] narrow band images obtained with VLT and presented in
  \citet{Liimets2018}. The contours in the right panel enhance the
  presence of the soft emission detected by {\it Chandra}.}
\label{fig:EPIC}
\end{center}
\end{figure*}

Its outstanding morphology has been also subject of exhaustive
studies.  R\,Aqr is surrounded by an exquisite filamentary and knotty
hourglass nebula as shown in optical wavelengths
(Fig.~\ref{fig:EPIC}).  The optical images show several bowl-shape
cavities open toward the north and south directions
\citep{Solf1985,Michalitsianos1988,Liimets2018}.  A close inspection
of the VLT [O\,{\sc iii}] image presented by \citet{Liimets2018}
reveals faint optical emission closing the southern lobe.  In
addition, images obtained with the 30~cm telescope at Terroux
Observatory reveal a faint closed lobe in the northern region
extending 2.8$'$ from R~Aqr. At the heart of this complex nebula,
there is a long-studied precessing jet with an S-shape morphology
\citep[see, e.g.,][]{Wallerstein1980,Paresce1994,Melnikov2018}. R~Aqr
is a complex system where all components seem interconnected (mass
loss of the primary, accretion onto the secondary, production of inner
intricate morphologies, etc.).  Hence, the dimming of the periodic
giant Mira-type star and the occurrence of the jets have been linked
to the periastron passage of the binary system \citep[see for
  example][]{Kafatos1982}.

The detection of high ionization features in the jets in the UV regime
\citep{Michalits1980} prompted several early X-ray studies.  X-rays
from R~Aqr were marginally detected with {\it Einstein} observations
\citep{Jura1984} and subsequently confirmed by {\it EXOSAT} and {\it
  ROSAT} observations \citep[see][]{Viotti1987,Hunsch1998}.  The X-ray
emission of R~Aqr remained unresolved until the advent of {\it
  Chandra}. \citet{Kellogg2001} used {\it Chandra} observations to
show that most of the X-ray emission is associated with the SS as well
as with clumps of the S-shaped jet.  The spectrum associated with the
SS shows the contribution from the Fe K$\alpha$ line at 6.4~keV, which
corroborated the accretion phenomena with material falling onto the WD
component.  \citet{Kellogg2001} showed that the X-ray-emitting clumps
in the S-shaped jet were shock-heated with plasma temperatures
$\gtrsim$10$^6$~K, similarly to what is found in extremely collimated
systems such as HH objects \citep{RK2019,Favata2002}, proto-planetary
nebulae \citep{Sahai2003}, nova-like systems
\citep{Toala2020,Montez2021} and other SSs
\citep[][]{Galloway2004,Stute2009,Stute2013}.

Subsequent X-ray studies performed by the same group demonstrated that
the X-ray-emitting clumps detected by {\it Chandra} exhibit
variability very likely due to the jet precession
\citep[e.g.,][]{Kellogg2007,Kellogg2009}. \citet{Kellogg2007} showed
that the spatial distribution of the X-ray-emitting gas changes within
a few years with an apparent projected velocity of 580$\times
(d/200$~pc)~km~s$^{-1}$, where $d$ is the distance to the object.  The
authors found that the clump located at the SW from the SS seemed to
fade between 2000 and 2004 and suggested that it might be due to its
adiabatic expansion and cooling, even though they predicted cooling
times longer than 800~yr.

In this letter we present the analysis of archival {\it XMM-Newton}
observations using reduction techniques for extended sources. We
report the detection of extended X-ray emission beyond the S-shaped
jets with a double-lobe morphology filling the space between the
hourglass nebular structures.  The observations and data preparation
are presented in Section~\ref{sec:obs}, while the results are
presented in Section~\ref{sec:results}. A closing discussion
addressing the origin of this extended emission and its spectral
properties in light with contemporaneous archival {\it Chandra}
observations is presented in Section~\ref{sec:discussion}.

\section{OBSERVATIONS and data preparation}
\label{sec:obs}

R Aqr was observed by {\it XMM-Newton} on 2005 June 30 using the
European Photon Imaging Camera (EPIC) (PI: E. Kellogg;
Obs. ID. 0304050101). The three EPIC cameras, namely MOS1, MOS2, and
pn, were used in the full frame mode with the medium optical blocking
filter for exposure times of 72.2, 72.2, and 70.6~ks,
respectively. The data were retrieved from the {\it XMM-Newton}
Science
Archive\footnote{\url{http://nxsa.esac.esa.int/nxsa-web/\#search}} and
processed with the Science Analysis Software \citep[SAS, version
  17.0;][]{Gabriel2004}, using the most recent calibrations available
by January 2022.

\begin{figure*}
\begin{center}
\includegraphics[width=0.9\linewidth]{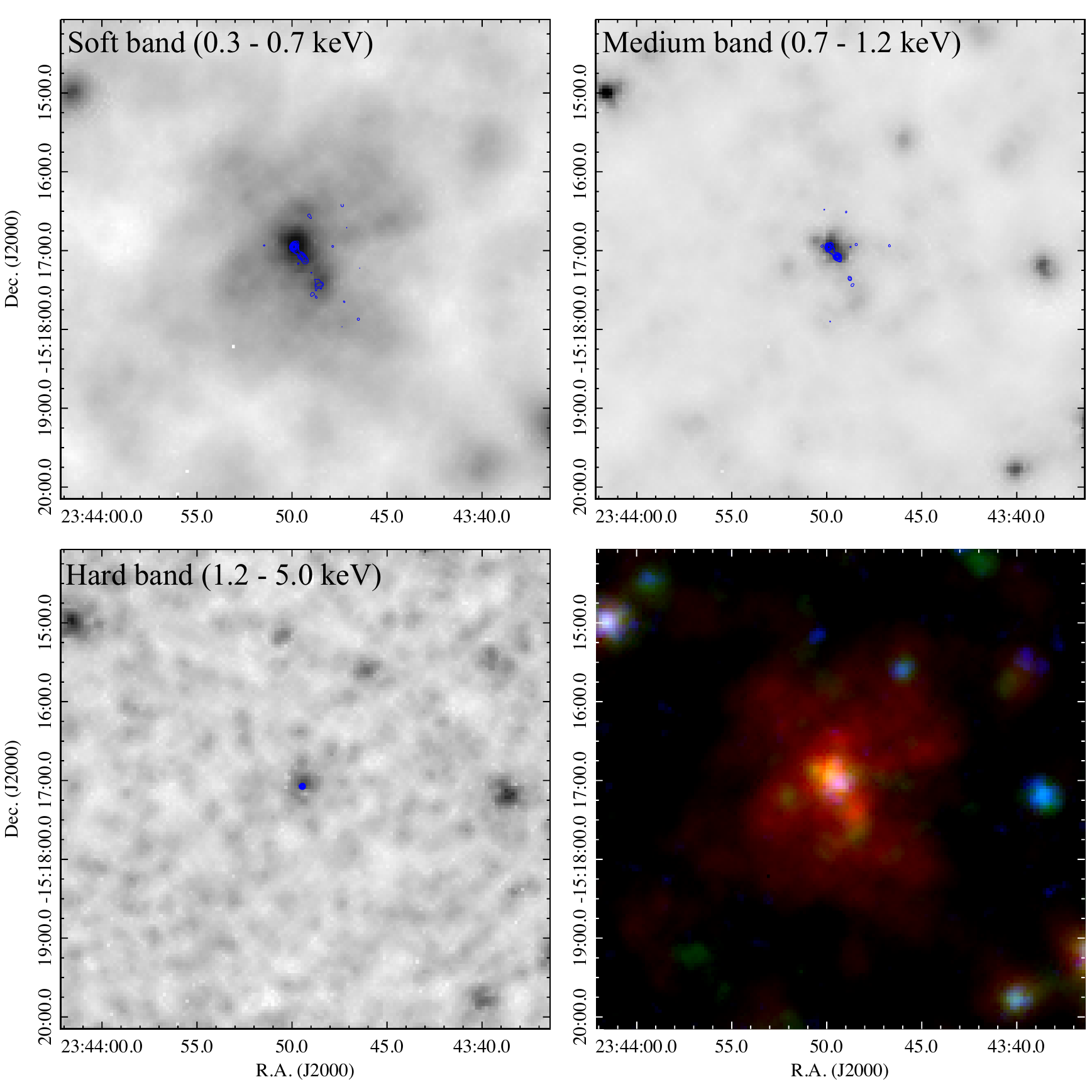}
\caption{{\it XMM-Newton} EPIC images of the X-ray emission of
  R~Aqr. Different panels show the soft (top left), medium (top right)
  and hard (bottom left) band images as well as a color-composite
  picture combining the three bands (bottom right).  The blue contours
  correspond to the X-ray emission detected by {\it Chandra} on each
  band.}
\label{fig:XMM}
\end{center}
\end{figure*}

\begin{figure*}
\begin{center}
\includegraphics[width=\linewidth]{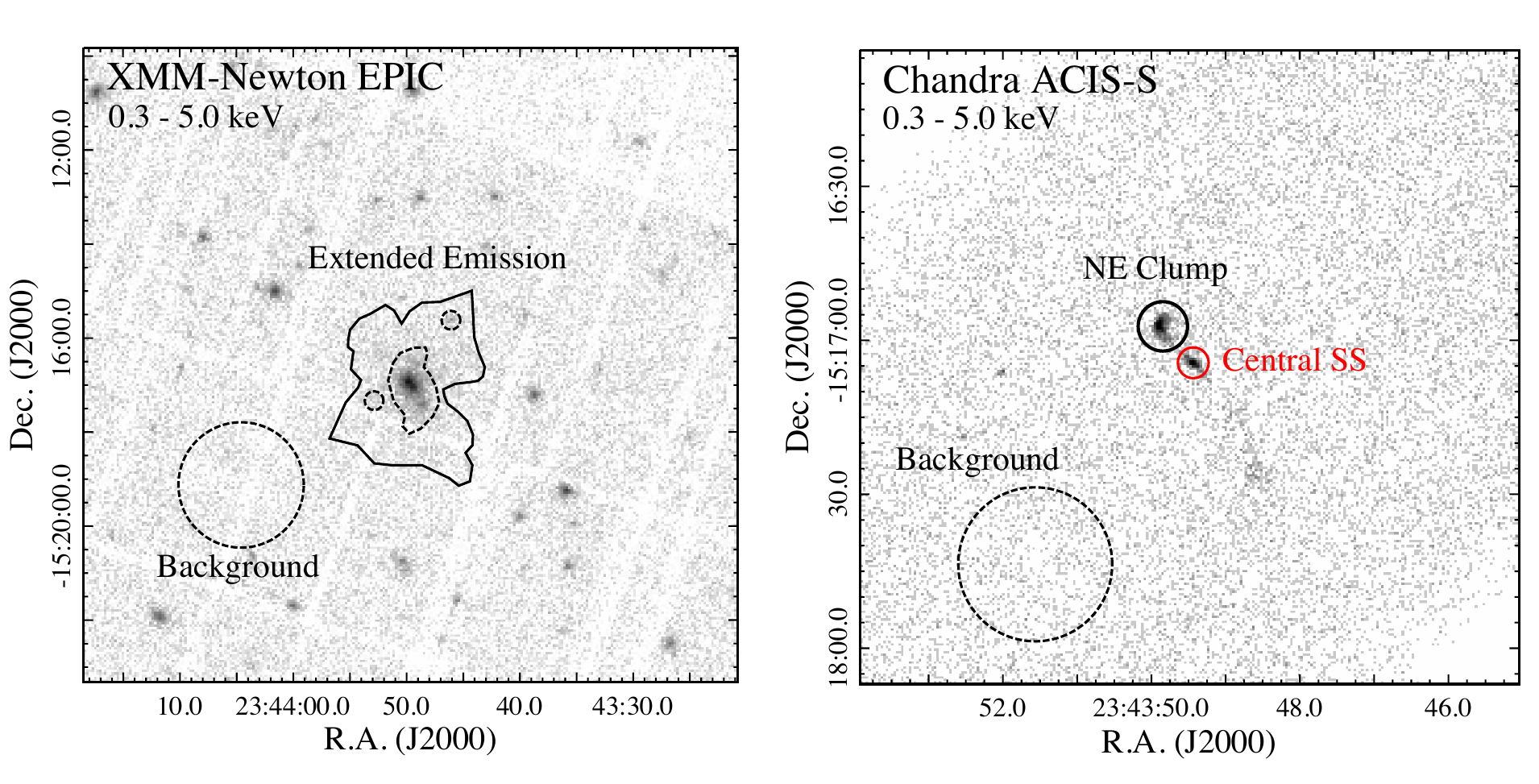}
\caption{{\it XMM-Newton} EPIC (left) and {\it Chandra} ACIS-S (right)
  event file images produced for the 0.3--5.0~keV energy range. The
  spectra extraction regions are shown with solid line regions whilst
  the background extraction regions are shown with dashed-line
  circular apertures. The dashed-line apertures inside the
    extended emission were excised from the spectra extraction
    procedure. Note that unlike \emph{XMM-Newton}, {\it Chandra}
  fully resolves the emission of the central SS from that of the NE
  clump.}
\label{fig:extraction}
\end{center}
\end{figure*}

To take advantage of the good quality of the {\it XMM-Newton}
observations of R Aqr, we analyzed the EPIC data in two different
ways. First, we started our analysis making use of the Extended Source
Analysis Software (ESAS) package, currently included as part of the
SAS. The data were processed following the cookbook for analysis of
extended objects and diffuse
background\footnote{\url{https://xmm-tools.cosmos.esa.int/external/sas/current/doc/esas/}}. The
ESAS tasks apply restrictive event selection criteria, but their
results leverage the presence of soft ($E<1.5$~keV) extended emission
\citep[see][]{Snowden2004,Snowden2008,Kuntz2008}.

After processing the data with the ESAS tasks, the final net exposure
times of the MOS1, MOS2, and pn cameras are 33.8, 39.1, and 20.4~ks
respectively. The processed files were used to create EPIC images in
the 0.3–0.7 keV (soft), 0.7–1.2 keV (medium), and 1.2–5.0 keV (hard)
energy bands.  Individual pn, MOS1, and MOS2 images were extracted,
corrected for exposure maps, and merged together.  The soft, medium,
and hard band images were adaptively smoothed using the ESAS task {\it
  adapt} requesting 30, 10, and 10~counts, respectively. The final
images are presented in Figure~\ref{fig:XMM} as well as a
color-composite X-ray image combining the three bands.

To study the physical properties of the extended X-ray emission
detected with {\it XMM-Newton}, we need to extract and model its
spectrum.  Since the ESAS event selection criteria are too restrictive
for this purpose, we reprocessed the EPIC data this time using the
{\it epproc} and {\it emproc} tasks.  To remove periods of high
background levels in the data, we produced light curves in the
10--12~keV energy range and excised time periods with count rates
higher than 0.2 and 0.45~counts~s$^{-1}$ for the MOS and pn cameras,
respectively. After excising bad time intervals, the event files have
net exposure times of 59.3, 60.0, and 35.8~ks for the MOS1, MOS2, and
pn, respectively.

For comparison and discussion we also retrieved {\it Chandra}
observations of R Aqr from the {\it Chandra} Data
Archive\footnote{\url{https://cda.harvard.edu/chaser/}}.  Several sets
of {\it Chandra} observations R\,Aqr are available, but, given its
X-ray variability, we choose to retrieve those observed at the same
epoch of the {\it XMM-Newton} EPIC data, i.e., those corresponding to
the Obs.\,ID.~5438 (PI: E.\,Kellogg) obtained on 2005 October 9 with a
total exposure time of 66.7~ks.  The data were obtained with the
Advanced CCD Imaging Spectrometer (ACIS)-S in the VFAINT mode and
R~Aqr was registered on the back-illuminated CCD~S3.  The {\it
  Chandra} data were processed with the Chandra Interactive Analysis
of Observations \citep[CIAO, version 4.7.4;][]{Fruscione2006}.

\section{Results}
\label{sec:results}

\subsection{Extended X-ray emission}

The {\it XMM-Newton} EPIC images produced with the ESAS tasks
unambiguosly reveal the presence of extended X-ray emission associated
with R~Aqr. Figure~\ref{fig:XMM} reveals the presence of X-ray
emission up to 2.2$'$ from the SS. In addition, Figure~\ref{fig:XMM}
shows that the bulk of the emission is emitted in the soft band. There
is some extended emission detected in the medium band distributed in
the inner $\sim$30$''$ very likely associated with the X-ray-bright
clumps resolved in the {\it Chandra} data \citep[see,
  e.g.,][]{Kellogg2007}. The hard X-ray emission is only associated
with the SS. This energy-dependent spatial distribution of the hot gas
can be appreciated in the color-composite picture presented in the
bottom-right panel of Figure~\ref{fig:XMM}.

\begin{figure*}
\begin{center}
\includegraphics[width=0.46\linewidth]{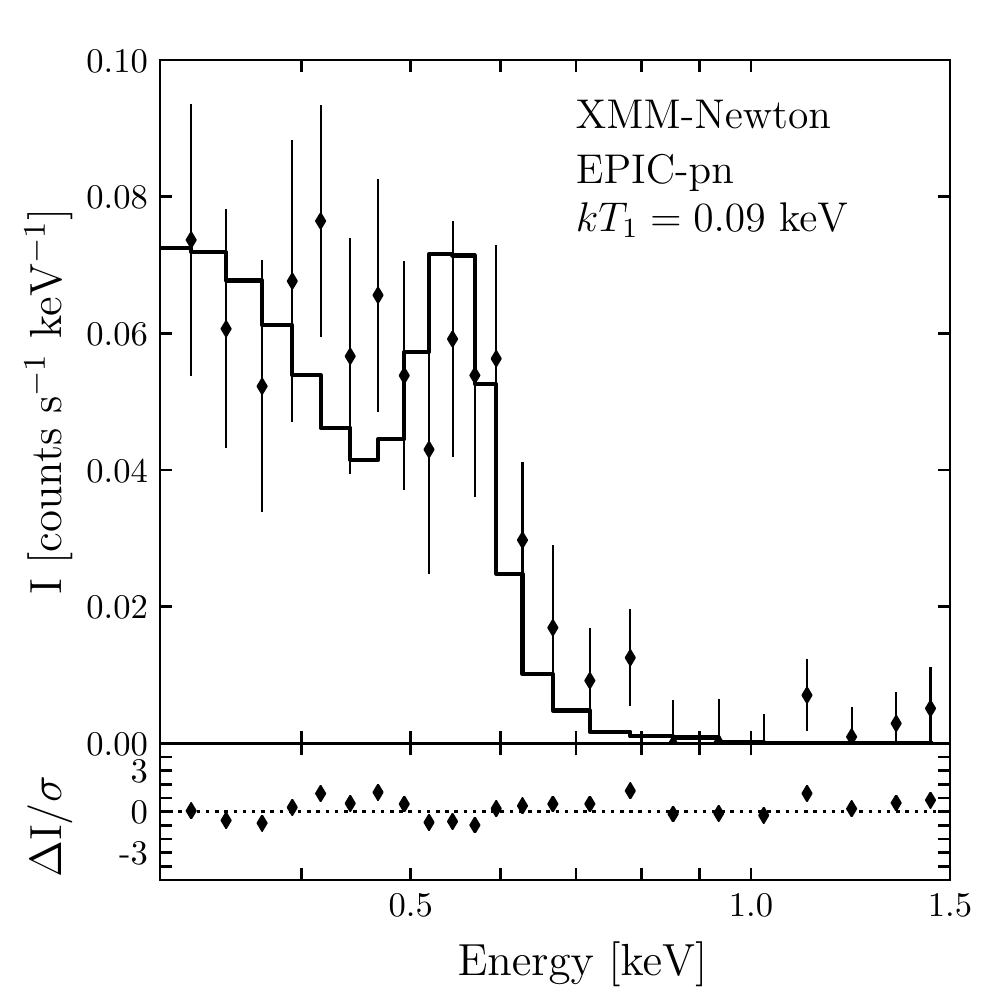}~
\includegraphics[width=0.46\linewidth]{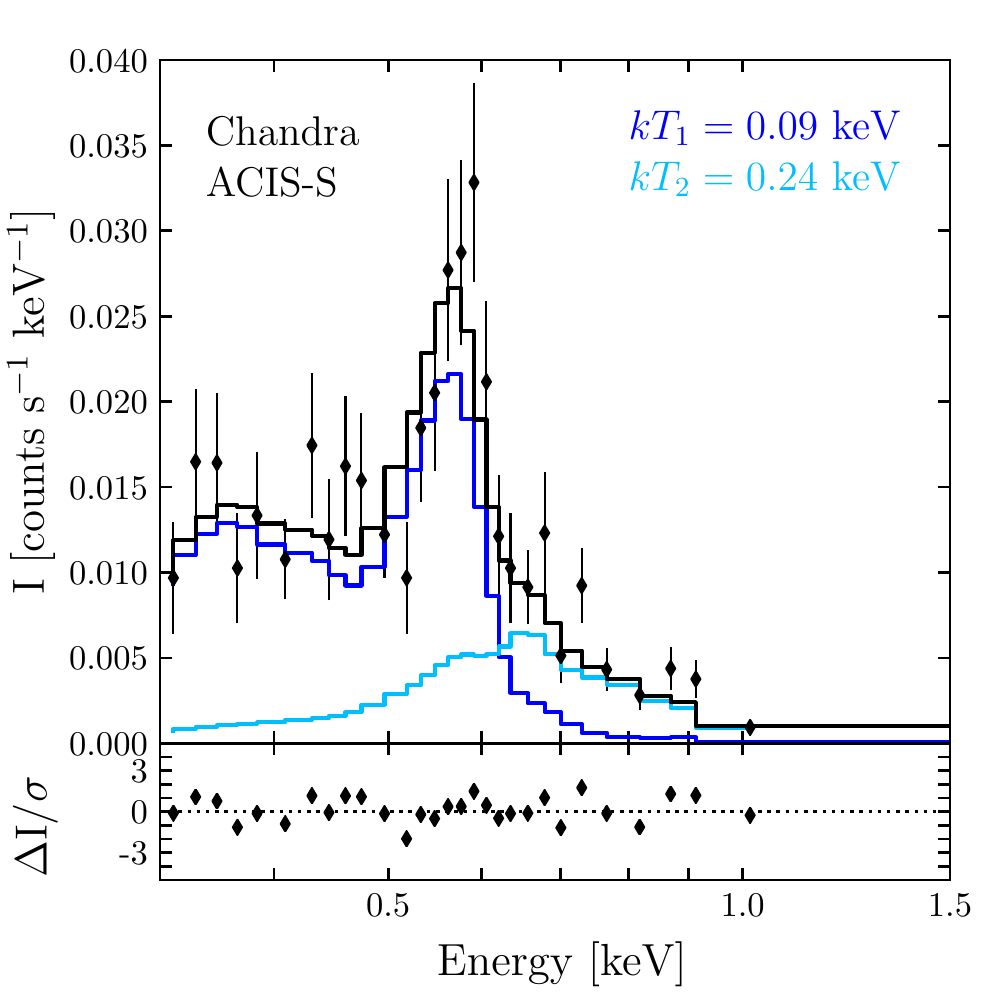}
\caption{(left) Background-subtracted {\it XMM-Newton} EPIC-pn
  spectrum of the large-scale diffuse X-ray emission of R\,Aqr (left)
  and {\it Chandra} ACIS-S spectrum of its bright NE clump (right).
  The black dots correspond to the observed spectra whilst the black
  histograms represent the best fit models. The blue and cyan histograms
    in the right panel illustrate individual single-temperature
  components. (bottom) Residuals of the spectral fit.}
\label{fig:spec}
\end{center}
\end{figure*}

Figure~\ref{fig:XMM} also suggests that the soft X-ray emission has a
bipolar morphology.  A pair of conical structures open up to the
northern and southern directions protruding from the central region.
This situation is further illustrated in Figure~\ref{fig:EPIC}-left
that compares the {\it XMM-Newton} soft X-ray emission with narrow
band images obtained with the VLT, making evident the match between
the distributions of the X-ray-emitting gas and the large-scale
optical nebular structures.

It is worth noting that the EPIC images does not resolve completely
the emission in the central region of R~Aqr, unlike the {\it Chandra}
observations that trace in detail the hot gas associated with the
bipolar jets of R~Aqr \citep[see][]{Kellogg2007,Nichols2007}.  On the
contrary, the \emph{Chandra} data present negligible contribution of
extended emission.  This situation is illustrated in
Figure~\ref{fig:EPIC}, which shows composite optical and X-ray
pictures using \emph{XMM-Newton} EPIC (left) and \emph{Chandra} ACIS-S
(right) data.  We confirm the different spatial-scales sampled by
\emph{XMM-Newton} and \emph{Chandra} in Figure~\ref{fig:extraction} by
comparing the event files of the EPIC and ACIS-S cameras.  The {\it
  Chandra} event files clearly resolve the contribution from the SS,
the NE clump, and some extended emission towards the SW (see
Fig.~\ref{fig:extraction} right panel), which can only be hinted in
the \emph{XMM-Newton} event files.  Contours of the soft, medium, and
hard emission detected by {\it Chandra} and generated with the CIAO
tasks {\it aconvolve} are also compared with the EPIC images in
Figure~\ref{fig:XMM}.

\subsection{Physical properties of the extended X-ray emission}

{\it XMM-Newton} spectra were extracted from the EPIC pn camera, given
its superior effective area compared with the MOS cameras.  The
spectrum of the extended emission was extracted from a region
encompassing the soft X-ray emission. We excluded the central region
which is dominated by the emission from the bipolar jet as well as a
pair of background point-like sources (see
Fig.~\ref{fig:extraction}). The background was extracted from a
circular region with a radius of 80$''$ towards the SE from R~Aqr (see
Fig.~\ref{fig:extraction} left panel).  The spectrum and associated
calibration matrices were created using the SAS tasks {\it eveselect},
{\it rmfgen} and {\it arfgen}. The resultant background-subtracted
EPIC pn spectrum of the extended emission is presented in the left
panel of Figure~\ref{fig:spec}. The net count rate detected from the
extended emission is 23.0$\pm$2.3~counts~ks$^{-1}$.

For comparison we also extracted the spectrum from the brightest X-ray
clump detected $\sim$10$''$ NE of R~Aqr and resolved in the {\it
  Chandra} ACIS-S data (see Fig.~\ref{fig:extraction} right panel).
Its extraction region corresponds to a 4$''$-radius circular aperture,
whilst the background was selected from a 15$''$-radius circular
aperture towards the SW.  The background-subtracted ACIS-S spectrum
and calibration matrices were produced using the CIAO task {\it
  specextract}. The resultant background-subtracted {\it Chandra}
ACIS-S spectrum of the NE clump is presented in the right panel of
Figure~\ref{fig:spec}. The NE clump has a net count rate of
6.9$\pm$0.3~counts~ks$^{-1}$.

Both spectra are extremely soft and resemble those presented and
analyzed by \citet{Kellogg2007}. They exhibit the presence of the
O\,{\sc vii} triplet at 0.58~keV and prominent emission at energies
below 0.5~keV.  The latter includes the very likely emission from
N\,{\sc vi} at 0.43~keV and the Ly$\alpha$ C\,{\sc vi} emission line
at 0.37~keV.  No significant emission is detected above 1.0~keV.

The spectral analysis was performed using XSPEC \citep{Arnaud1996}. We
started by fitting single plasma emission models to the data, which we
note it was enough for the EPIC spectrum, but according to
\citet{Kellogg2007} the {\it Chandra} spectrum requires
two-temperature {\it apec} optically-thin emission plasma models.  The
absorption was included with the {\it tbabs} model \citep{Wilms2000}
and all spectra were fit adopting solar abundances from
\citet{Lodders2003}.

\begin{table}
  \begin{center}
\caption{Spectral fits of the diffuse X-ray emission in R~Aqr.}
\footnotesize
\setlength{\tabcolsep}{0.9\tabcolsep}    
\begin{tabular}{llcc}
\hline
Parameter        &                       &    Extended Emission           & NE Clump \\
                 &                       &   (EPIC pn)                    & (ACIS-S) \\
\hline
$\chi^2$         &                       &       1.00                     & 1.04                   \\
$N_\mathrm{H}$   &   [$10^{20}$~cm$^{-2}$] & 4.0 (fixed)                  & $8.0^{+11}_{-5.0}$   \\
$kT_{1}$         &   [keV]               & $0.09^{+0.01}_{-0.02}$.        & $0.09^{+0.02}_{-0.03}$ \\
$kT_{2}$         &   [keV]               & \dots                          & $0.24^{+0.09}_{-0.05}$ \\
$A_{1}$          &   [cm$^{-5}$]         & 1.6$\times10^{-4}$             & 4.3$\times10^{-4}$    \\
$A_{2}$          &   [cm$^{-5}$]          & \dots                         & 1.2$\times10^{-5}$    \\
$f_\mathrm{X}$ & [erg~cm$^{-2}$~s$^{-1}$] & (4.4$\pm$1.5)$\times$10$^{-14}$ & (4.8$\pm$0.8)$\times$10$^{-14}$ \\
$F_\mathrm{X}$  & [erg~cm$^{-2}$~s$^{-1}$]&(7.7$\pm$2.6)$\times$10$^{-14}$  & (1.5$\pm$0.2)$\times$10$^{-13}$ \\
$L_\mathrm{X}^\dagger$  & [erg~s$^{-1}$]  & (1.4$\pm$0.5)$\times$10$^{30}$  & (2.7$\pm$0.4)$\times$10$^{30}$ \\
\hline
\end{tabular}
\label{tab:parameters}
\end{center}
  \vspace{-3.5mm}
  \footnotesize{$^{\dagger}$Adopting a distance of 385~pc.}
\end{table}

We started by fitting the {\it Chandra} ACIS-S spectrum of the NE
clump. The best fit model ($\chi^{2}$=1.04) resulted in plasma
temperatures of $kT_{1}=0.09^{+0.02}_{-0.03}$~keV and
$kT_{2}=0.24^{+0.09}_{-0.05}$~keV with normalization
parameters\footnote{The normalization parameter can be estimated as $A
  \approx 10^{-14} \int n_\mathrm{e}^{2} dV / 4 \pi d^{2}$, where
  $n_\mathrm{e}$ is the electron number density, $V$ is the volume of
  the X-ray-emitting region and $d$ is the distance to the object.} of
$A_{1}=4.3\times10^{-4}$~cm$^{-5}$ and
$A_{2}=1.2\times10^{-5}$~cm$^{-5}$, respectively. The absorption
column density of the model was
$N_\mathrm{H}=(8^{+11}_{-5})\times10^{20}$~cm$^{-2}$, consistent with
the extinction estimate of $A_\mathrm{V}$=1--2~mag by
\citet{Bujarrabal2021} towards R~Aqr. The total observed flux in the
0.3--1.5~keV energy range is
$f_\mathrm{X}=(4.8\pm0.8)\times10^{-14}$~erg~cm$^{-2}$~s$^{-1}$ which
corresponds to an intrinsic flux of
$F_\mathrm{X}=(1.5\pm0.2)\times10^{-13}$~erg~cm$^{-2}$~s$^{-1}$ and an
X-ray luminosity\footnote{Here we adopt the geometric {\it Gaia}
    distance, but we note that smaller distance estimates have been
    reported in the literature. The more recent is that of
    \citet{Liimets2018} which obtained a distance of $\sim$180~pc
    using the expansion parallax method.} of
$L_\mathrm{X}=(2.7\pm0.4)\times10^{30}$~erg~s$^{-1}$. We note that the
softer component contributes to 85\% of the total X-ray
emission. Finally, we estimated an electron density of
$n_\mathrm{e}\approx40$~cm$^{-3}$ by adopting an averaged radius of
4$''$ for the NE clump. The model is compared with the observed {\it
  Chandra} ACIS-S spectrum in the right panel of Figure~\ref{fig:spec}
and the details are listed in Table~\ref{tab:parameters}.

The spectral fit of the EPIC pn spectrum of the extended emission
resulted in very similar properties as those of the NE clump. However,
we note that we had to fix the absorption column density, because when
left as a free parameter it converged to unphysical values.  The
$N_\mathrm{H}$ in the extended regions away from R~Aqr should have
less extinction than that of the central regions. The HEASARC
$N_\mathrm{H}$ column density
webpage\footnote{\url{https://heasarc.gsfc.nasa.gov/cgi-bin/Tools/w3nh/w3nh.pl}}
estimates that in the vicinity of R~Aqr this is
$\sim2\times10^{20}$~cm$^{-2}$. Thus, we decided to fix the absorption
column density to an intermediate value of
$N_\mathrm{H}=4\times10^{20}$~cm$^{-2}$. With this, the best fit model
($\chi^{2}=$1.00) has a dominant plasma temperature of
$kT_{1}=0.09^{+0.01}_{-0.02}$~keV with normalization parameter of
$A_{1}=1.6\times10^{-4}$~cm$^{-5}$. The total observed
flux in the 0.3--1.5~keV energy range is
$f_\mathrm{X}=(4.4\pm1.5)\times10^{-14}$~erg~cm$^{-2}$~s$^{-1}$ with
an intrinsic flux of
$F_\mathrm{X}=(7.7\pm2.6)\times10^{-14}$~erg~cm$^{-2}$~s$^{-1}$, which
corresponds to an X-ray luminosity of
$L_\mathrm{X}=(1.4\pm0.5)\times10^{30}$~erg~s$^{-1}$.
The estimated electron density for the most extended emission of 
$n_\mathrm{e}\approx$0.5~cm$^{-3}$ adopting an averaged angular
radius of 2$'$. The best fit is compared to the EPIC pn spectrum in
presented in the left panel of Figure~\ref{fig:spec} and the model
details are also listed in Table~\ref{tab:parameters}.

We note that the observed surface brightness of the extended emission
detected by {\it XMM-Newton} is a few times
10$^{-17}$~erg~s$^{-1}$~cm$^{-2}$~arcsec$^{-2}$. This value is just at
the detection limit of the ACIS-S instrument, thus explaining why it
was not detected by {\it Chandra}.

Other models to the X-ray spectra were attempted.  For example, we
also tried plasma emission models with variable N abundance in order
to fit the spectral feature at 0.43~keV, but such models did not
improve the fit.  Models adopting a power law distribution for the
hardest component were also attempted.  Such models resulted in
good-quality fits ($\chi^{2} \approx 1$) too, suggesting that the
physical origin of the second component can not be firmly established.
Lastly, we note that other models taking into account non
  equlibrium thermal components or non equlibrium collisional model
  are difficult to constrain given the relative poor-quality of the
  spectra \citep[see][]{Kellogg2007}.

\section{Discussion and concluding remarks} 
\label{sec:discussion}

The analysis of the {\it XMM-Newton} EPIC data with the ESAS tasks has
revealed the presence of extended emission surrounding R~Aqr in the
soft (0.3--0.7~keV) X-ray band.  The diffuse emission detected by EPIC
seems to fill the gaps in between the nebular hourglass structures,
extending to distances more than 2$'$ towards the northern and
southern regions from R~Aqr (see, e.g., Fig.~\ref{fig:EPIC} left
panel).  The extension of the diffuse X-ray emission is consistent
with the results discussed in \citet{Bujarrabal2021} where the orbital
plane of the SS is seen almost edge-on and most of the dense gas is
distributed in the equatorial plane, with lower density along the N-S
directions that facilitates the expansion of the hot gas.

We found that the spectrum of the extended X-ray emission detected by
{\it XMM-Newton} EPIC shares a very cool component with the spectrum
of the NE clump associated with the S-shaped jet resolved by {\it
  Chandra} ACIS-S. This suggests that the jet is an ongoing structure
that keeps feeding the extended, X-ray-emitting hot bubble. The
cooling time of such hot but low density gas is long
\citep[see][]{Kellogg2007} and, accordingly, no significant reduction
of the temperature is observed between the jet and the most extended
X-ray emission. The estimated $n_\mathrm{e}$ of the most extended gas
is almost 80 times lower than that associated with the NE clump,
implying a significantly smaller differential emission measure for the
former. Thus, the gas expands adiabatically and cools down.

The nature of the harder component required in the spectral analysis
of the X-ray-emitting jet is still unknown. This seems to be only
spatially correlated with the collimated jet and it is not surprising
that magnetic phenomena are responsible for the shaping
\citep{Burgarella1992,Melnikov2018}. However, the search for non
thermal emission in radio bands has been challenging
\citep{Hollis1985,Bujarrabal2018}. Further analysis combining
multi-epoch {\it Chandra} data will help shed light on the physical
properties of the X-ray-emitting gas associated with jets (R.\,Montez
Jr private communication).

The jet appears to be collimated, its S-shape is clearly indicative
of precession. The precession angle is wide, $\approx50^{\circ}$, as
suggested by high-resolution {\it Hubble Space Telescope} images
\citep{Melnikov2018}, and actually theoretical predictions state that
opening angles larger than 40$^{\circ}$ produce nebulae with pairs of
inflated bubbles \citep{Soker2004}. We therefore suggest that
blister-like structures at the tip of the precessing jet get disrupted
and feed hot gas into the most extended hot bubble.  The
creation-destruction of the blister-like structures seems to be
relatively fast given the evident expansion of the X-ray-emitting
clumps resolved by {\it Chandra} \citep[e.g.,][]{Kellogg2007}.  It is
possible that the high pressure of the hot gas helped inflating the
most extended lobe detected in optical wavelengths \citep[see figure~2
  of][]{Liimets2018}.

We also suggest that the continuous production and disruption of
blister-like structures has created the bowl-like structures opening
towards the northern and southern regions of R~Aqr.  Disk formation
and jet launching has been studied through numerical simulations of
binary systems in great detail \citep[see, e.g.,][and references
  therein]{LopezCamara2020,Sheikhnezami2018,Waters2018}, but the {\it
  XMM-Newton} EPIC observations analyzed here leverage the role of
accretion process onto a compact object, a WD in this case, on the
formation of hot bubbles and large-scale nebulae.  This process might
be similar, but in an evidently smaller scale, to the creation of hot
bubbles produced by active galaxies hosting supermassive black holes
\citep[see, e.g., the case of Cen~A;][and references
  therein]{Krol2020,Hardcastle2003}.

Several theoretical works have proposed the jet feedback mechanism
(JFM) as the physics behind the production of inflated bipolar
cavities in a wide variety of astronomical phenomena.  The JFM
successfully explains a diverse variety of objects regardless of
differences of several orders of magnitude in size, energy injection,
mass and timescales \citep{Soker2016}, from young stellar objects
through planetary nebulae to galaxy clusters
\citep[e.g.,][]{Cliffe1995,Soker2002,Soker2004,Soker2016}.

Deeper and higher-quality observations, as those provided by {\it
  eROSITA} and future missions such as {\it Athena}, might be able to
disclose more extended X-ray emission with lower surface brightness of
X-ray-emitting gas in R~Aqr. A complete mapping of the diffuse X-ray
hot bubble will help us study the feedback from SSs to create fast
evolving nebulae around them.  \\

\noindent 
The authos thank the anonymous referee for comments and suggestions
that improved the analysis presented here. The authors are thankful
to Tiina Liimets for providing the VLT narrow band images of
R\,Aqr. They also acknowledge Rodolfo Montez Jr for fruitful
discussion during the preparation of the manuscript. JAT acknowledges
support by the Marcos Moshinsky Foundation (Mexico) and UNAM PAPIIT
project IA101622 (Mexico). LS acknowledges support by UNAM PAPIIT
project IN110122 (Mexico).  MAG acknowledges support of grant PGC
2018-102184-B-I00 of the Ministerio de Educación, Innovación y
Universidades cofunded with FEDER funds. GR-L acknowledges support
from CONACyT grant 263373. YHC acknowledges the grant MOST
110-2112-M-001-020 from the Ministry of Science and Technology
(Taiwan).  Based on observations obtained with {\it XMM-Newton}, an
ESA science mission with instruments and contributions directly funded
by ESA Member States and NASA. This research has made use of data
obtained from the {\it Chandra} Data Archive and software provided by
the {\it Chandra} X-ray Center (CXC) in the application package CIAO.
This work has made extensive use of the NASA's Astrophysics Data System.



\software{SAS \citep{Gabriel2004}, XSPEC \citep{Arnaud1996}}, CIAO \citep{Fruscione2006}
\facilities{{\it Chandra}, {\it XMM-Newton}, Very Large Telescope (VLT)}



\FloatBarrier

\end{document}